\documentclass[]{aa}
\usepackage{natbib}
\usepackage{graphics}
\usepackage{txfonts}

\begin{document}

\title{N-body simulations in reconstruction of the kinematics of young stars in the Galaxy}
\titlerunning{N-body simulations of the Galaxy}
\author{P. Rautiainen\inst{1}\thanks{E-mail: pertti.rautiainen@oulu.fi}  \and A.M.
Mel'nik\inst{2}}

\institute{Department of Physics/Astronomy Division,
University of Oulu, P.O. Box 3000,\\ FIN-90014 Oulun yliopisto,
Finland \\ \and Sternberg Astronomical Institute, 13,
Universitetskii pr., Moscow, 119992, Russia}

\date{Received 2010}

\abstract {} 
{We try to determine the Galactic structure by comparing the
 observed and modeled velocities of OB-associations in the 3 kpc solar
neighborhood.}
{We made N-body simulations with a rotating stellar bar. The
  galactic disk in our model includes gas and stellar subsystems. The
  velocities of gas particles averaged over large time intervals
  ($\sim 8$ bar rotation periods) are compared with the observed
  velocities of the OB-associations.}
{Our models reproduce the directions of the radial and azimuthal
components of the observed residual velocities  in
 the Perseus and Sagittarius regions and in the Local system. The
  mean difference between the model and observed
  velocities is $\Delta V=3.3$ km s$^{-1}$. The optimal value of the
  solar position angle $\theta_b$ providing the best agreement between
  the model and observed velocities is $\theta_b=45\pm5^\circ$, in
  good accordance with several recent estimates. The self-gravitating
  stellar subsystem forms a bar, an outer ring of subclass $R_1$, and
  slower spiral modes. Their combined gravitational perturbation leads
  to time-dependent morphology in the gas subsystem, which forms outer
  rings with elements of the $R_1$- and $R_2$-morphology. The success
  of N-body simulations in the Local System is likely due to the
  gravity of the stellar $R_1$-ring, which is omitted in models with
  analytical bars.} {}

\keywords {Galaxy: structure: Galaxy: kinematics and
dynamics}

\maketitle

\section{Introduction}

The consensus since the 1990s has been that the Milky Way is a barred
galaxy \citep[see, e.g.][]{blitz1991,blitz1993}. The estimate for the
size of the large-scale bar has grown from initial $R_{bar} \approx
2-3$ kpc to current estimates $R_{bar}=3-5$ kpc
\citep{habing2006,cabrera-lavers2007,cabrera-lavers2008}. The
  position angle of the bar is thought to be in the range $15^\circ -
  45^\circ$
  \citep[][]{blitz1993,kuijken1996,weiner1999,benjamin2005,englmaier2006,cabrera-lavers2007,minchev2009}. The
  differences in the position angle estimates may indicate that the
  innermost structure is actually a triaxial bulge
  \citep{cabrera-lavers2008}. On the other hand, this ambiguity may
be partly caused by our unfavorable viewing angle near the disk plane,
which also hinders study of other aspects of Galactic morphology.

The suggested configurations for the spiral morphology of the Galaxy
include models or sketches containing from two to six spiral arms
\citep[see e.g.][and references therein]{vallee2005,vallee2008}. A
case has also been suggested where a two-armed structure dominates in
the old stellar population, whereas the gas and young stellar
population exhibits a four-armed structure
\citep{lepine2001,churchwell2009}. In addition to spiral arms, there
may be an inner ring or pseudoring surrounding the bar, which
manifests itself as the so-called 3-kpc arm(s)
\citep{dame2008,churchwell2009}. Also, speculations about a nuclear
ring with a major axis of about 1.5 kpc have been made
\citep{rodriguez-fernandez2008}. Different kinds of rings -- nuclear
rings, inner rings and outer rings -- are often seen in the disks of
spiral galaxies, especially if there is also a large-scale bar
\citep{buta1996}. Thus, the presence of an outer ring in the Galaxy
may also be considered plausible \citep{kalnajs1991}.

Since the outer rings have an elliptic form, the broken outer rings
(pseudorings) resemble two tightly wound spiral arms. Nevertheless
their connection with the density-wave spiral arms is not very obvious
because their formation does not need the spiral-shaped perturbation in
the stellar disk. The main ingredient for their formation is a
rotating bar. Both test particle simulations \citep{schwarz1981,
  byrd1994,bagley2009} with an analytical bar and N-body simulations
\citep{rautiainen1999,rautiainen2000}, where the bar forms in the disk
by instability, show that the outer rings and pseudorings are typically
located in the region of the Outer Lindblad Resonance (OLR). Two main
classes of the outer rings and pseudorings have been identified: the
$R_1$-rings and $R'_1$-pseudorings elongated perpendicular to the bar
and the $R_2$-rings and $R'_2$-pseudorings elongated parallel to the
bar.  In addition, there is a combined morphological type $R_1R_2'$
that shows elements of both classes \citep{buta1986, buta1991,
  buta1995, buta1996, buta2007}.

\citet{schwarz1981} connected two main types of the outer rings with
two main families of periodic orbits existing near the OLR of the bar
\citep{contopoulos1980,contopoulos1989}. The stability of orbits
enables gas clouds to follow them for a long time period. The
$R_1$-rings are supported by $x_1(2)$-orbits \citep[using the
  nomenclature of][]{contopoulos1989} lying inside the OLR and
elongated perpendicular to the bar, while the $R_2$-rings are
supported by $x_1(1)$-orbits situated a bit outside the OLR and
elongated along the bar. There is also another conception of the ring
formation. \citet{romerogomez2007} show that Lyapunov periodic orbits
around $L_1$ and $L_2$ equilibrium points can lead to the formation of
the spiral arms and the outer rings. They associate the spiral arms
emanating from the bar's tips with the unstable manifolds of Lyapunov
orbits. This approach can be useful for explaining of the motion of
gas particles as well \citep{athanassoula2009}.

Besides the bar the galactic disks often contain spiral arms, which
modify the shape of the gravitational perturbation. In the simplest
case, the pattern speeds of the bar and spiral arms are the same. In
many studies this assumption has been used for constructing the
gravitational potential from near-IR observations (which represent the
old stellar population better than the visual wavelengths). Several
galaxies with outer rings have been modeled by this method, and
findings are in good accordance with studies made by using analytical
bars: the outer rings tend to be located near the OLR
\citep{salo1999}, although in some cases they can be completely
confined within the outer 4/1-resonance, \citep{treuthardt2008}.

A real galactic disk provides further complications, which can be
studied by N-body models, where the bars and spiral arms are made of
self-gravitating particles. In particular, there can often be one or
more modes rotating more slowly than the bar
\citep{sellwood1988,masset1997,rautiainen1999}. Even if there is an
apparent connection between the ends of the bar and the spiral arms,
it is no guarantee that the pattern speeds are equal -- the break
between the components may be seen only for a short time before the
connection reappears \citep[see Fig. 2 in][]{sellwood1988}. Sometimes
the bar mode can contain a considerable spiral part that forms the
observed spiral, together with the slower modes
\citep{rautiainen1999}. The multiple modes can also introduce cyclic
or semi-cyclic variations in the outer spiral morphology: outer rings
of different types can appear and disappear temporarily
\citep{rautiainen2000}.

In \citet{melnikrautiainen2009} (hereafter Paper I), we considered
models with analytical bars. In this case the motion of gas particles
is determined only by the bar. We found that the resonance between the
epicyclic motion and the orbital motion creates systematical
noncircular motions that depend on the position angle of a point
with respect to the bar elongation and on the class of the outer
ring. The resonance kinematics typical of the outer ring
of subclass $R_1R_2'$ reproduces the observed velocities in the
Perseus and Sagittarius regions well.

In paper I we also suggested that the two-component outer ring could
be misinterpreted as a four-armed spiral. In some galaxies
with the combined $R_1R_2'$-morphology, the $R_1$-component can also be seen
in the near infrared, but the $R_2$-component is usually
prominent only in blue \citep{byrd1994}. This could explain the
  ambiguity of the number of spiral arms in the Galaxy. N-body
simulations confirm that the $R'_1$-rings can be forming in the
self-gravitating stellar subsystem, while the $R_2'$-rings usually
exist only in the gas component \citep{rautiainen2000}. 

In the present paper we study the effect of multiple modes and their
influence on the kinematics and distribution of gas particles.  We
construct N-body models to study the influence of self-gravity
in the stellar component on the kinematics of gas particles.  We
compare the model velocities of gas particles with the observed velocities
of OB-associations in the neighborhood 3 kpc from the Sun.

This paper has the following structure. Observational data are considered
in Sect. 2. Section 3 is devoted to models and describes the
essential model parameters, the evolution of the stellar and gas
components: formation of the bar and the interplay between the bar and
slower spiral modes. In Sect. 3 we also analyze the general features
of the gas morphology. Section 4 is devoted to the comparison between
the observed and modeled kinematics. Both momentary and average
velocities of gas particles are considered. The influence of the bar
position angle $\theta_b$ on the model velocities is also investigated
in Sect. 4, as are the evolutionary aspects of kinematics.
Section 5 consists of conclusions and discussion.

\section{Observational data}

\begin{figure*}
\resizebox{\hsize}{!}{\includegraphics{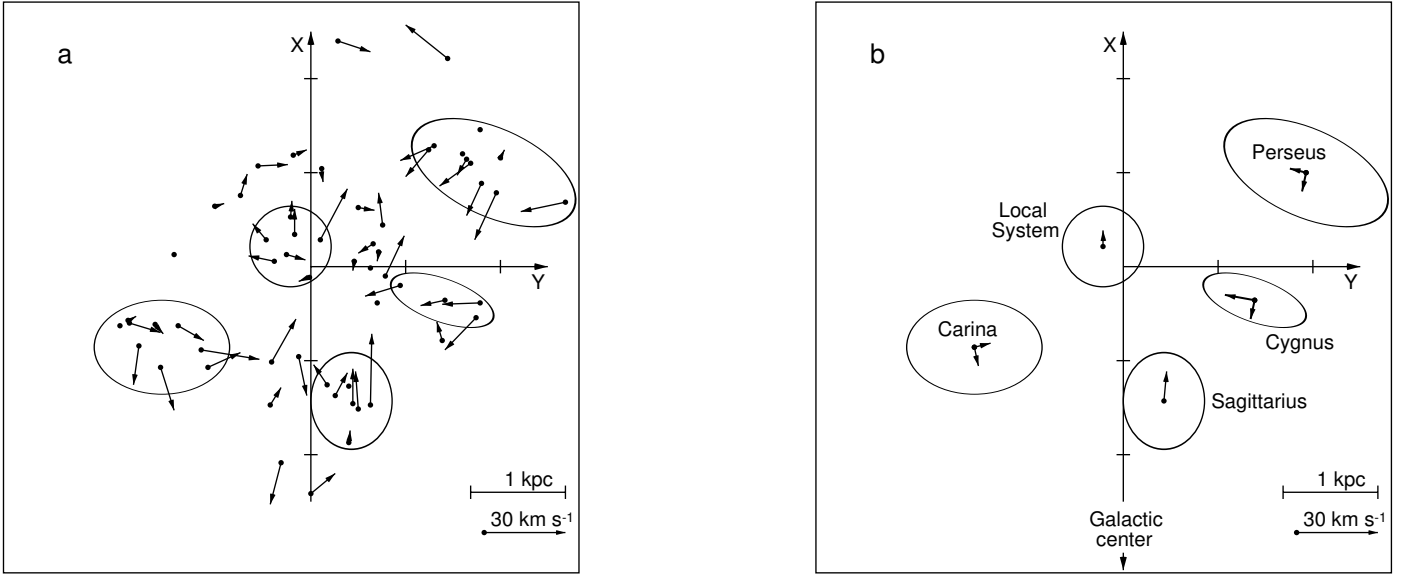}}
\caption{(a) The residual velocities of OB-associations projected
on to the galactic plane. It also shows the grouping of
OB-associations into regions of intense star formation. (b) The
mean  $V_R$- and $V_\theta$- velocities of OB-associations in the
stellar-gas complexes. The X-axis is directed away from the
galactic center, and the Y-axis is in the direction of the
galactic rotation. The Sun is at the origin.} \label{complexes}
\end{figure*}

We have compared the mean residual velocities of OB-associations in the
regions of  intense star formation with those of gas particles in
our models. These regions practically
coincide with the stellar-gas complexes identified by
\citet{efremov1988}. The residual velocities characterize the
non-circular motions in the galactic disks. They are calculated
as differences between the observed heliocentric velocities
(corrected for the motion to the apex) and the velocities due to
the circular rotation law. We used the list of OB-associations by
\citet{BlahaHumphreys1989}, the line-of-sight velocities
\citep{barbierbrossat2000}, and proper motions
\citep{hipparcos1997, vanleeuwen2007} to calculate their median
velocities along the galactic radius-vector, $V_R$, and in the
azimuthal direction, $V_\theta$. Figure ~\ref{complexes} shows the
residual velocities of OB-associations in the regions of intense
star formation. It also indicates the grouping of OB-associations
into stellar-gas complexes. For each complex we calculated the
mean residual velocities of OB-associations, which are listed in
Table~\ref{observations}. Positive radial residual velocities
$V_R$ are directed away from the Galactic center, and the
positive azimuthal residual velocities $V_\theta$ are  in the
sense of Galactic rotation. Table~\ref{observations} also
contains the rms errors of the mean velocities, the mean
Galactocentric distances $R$ of OB-associations in the complexes,
the corresponding intervals of galactic longitudes $l$ and
heliocentric distances $r$, and names of OB-associations the
region includes \citep[see also][]{melnikdambis2009}.

\begin{table*}
  \caption{Observed  residual velocities of OB-associations in the stellar-gas complexes}
  \begin{tabular}{lcccccl}
  \hline
  Region& {\it R} & $V_{R\mbox{ obs}}$ & $V_{\theta\mbox{ obs}}$  & {\it l}  & {\it r} & Associations \\
    &  kpc& km s$^{-1}$ & km s$^{-1}$  & deg. & kpc &  \\
  \\[-7pt]\hline\\[-7pt]
  Sagittarius & 5.6 & $+9.9\pm2.4$ & $-1.0\pm1.9$ & 8--23  & 1.3--1.9 & Sgr OB1, OB7, OB4, Ser OB1, OB2, \\
  & & & & & &  Sct OB2, OB3;\\
  Carina & 6.5 & $-5.8\pm3.3$ & $+4.7\pm2.2$ & 286--315  & 1.5--2.1 & Car OB1, OB2, Cru OB1, Cen OB1,\\
  & & & & & &   Coll 228, Tr 16, Hogg 16, NGC 3766, 5606;\\
  Cygnus & 6.9 & $-5.0\pm2.6$ & $-10.4\pm1.4$ & 73--78  & 1.0--1.8 & Cyg OB1, OB3, OB8, OB9; \\
  Local System & 7.4 & $+5.3\pm2.8$ & $+0.6\pm2.5$ & 0--360  & 0.1--0.6 & Per OB2, Mon OB1, Ori OB1, Vela OB2, \\
  & & & & & &   Coll 121, 140, Sco OB2; \\
  Perseus & 8.4 & $-6.7\pm3.0$ & $-5.9\pm1.5$ & 104--135  & 1.8--2.8 & Per OB1, NGC 457, Cas OB8, OB7, OB6, \\
  & & & & & &  OB5, OB4, OB2, OB1, Cep OB1;\\
  \hline
\end{tabular}
\label{observations}
\end{table*}

The Galactic rotation curve derived from an analysis of the
kinematics of OB-associations is nearly flat in the 3-kpc solar
neighborhood and corresponds to the linear velocity at the solar
distance of $\Theta_0=220$ km s$^{-1}$
\citep{melnik2001,melnikdambis2009}. The nearly flat form of
the Galactic rotation curve  was found in many other studies
\citep{burton1978, clemens1985, brand1993, pont1994, dambis1995,
russeil2003, bobylev2007}.

We adopted the Galactocentric distance of the Sun to be $R_0=7.5$
kpc \citep[][and other papers]{rastorguev1994, dambis1995,
glushkova1998}, which is consistent with the so-called short
distance scale for classical Cepheids \citep{berdnikov2000}.

\section{Models}

\subsection{The model parameters}

We made several N-body models, which satisfy ``broad observational
constraints'': the rotation curve is essentially flat and the size of
the bar is acceptable. From these models we have chosen our
best-fitting case, which we describe here in more detail.

\begin{figure*}
\centering
\includegraphics{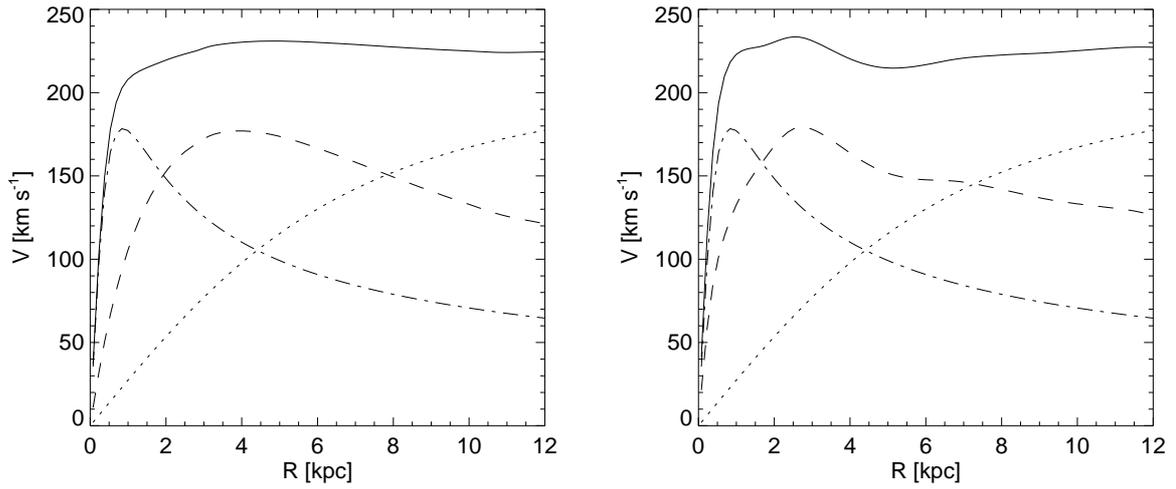}
\caption{The rotation curve (solid line) of the N-body model at $T=0$
  (left) and at $T=55$ (right). The contributions from the bulge
  (dash-dotted line), disk (dashed line) and halo (dotted line) are
  also indicated.} \label{nbody_rcurve}
\end{figure*}

The rotation curve of our best-fitting model is illustrated in
Fig.~\ref{nbody_rcurve}. In the beginning, the rotation curve is
slightly falling in the solar neighborhood, but the mass rearrangement
in the disk during the bar formation makes it rise slightly. We scaled
the simulation units to correspond to our preferred values of the
solar distance from the Galactic center and the local circular
velocity. This also gives the scales for masses and time
units. However, in the following discussion we will use simulation
time units, one corresponding to approximately 100 million years, and
the full length of the simulation is 6 Gyrs.

The bulge and halo components are analytical, whereas the stellar disk is
self-gravitating.  The bulge is represented by a Plummer sphere, mass
$M_{bulge}=1.17 \times 10^{10} \ M_\odot$, and scale length
$R_{bulge}=0.61$ kpc. The dark halo was included as a component giving
a halo rotation curve of form

\begin{equation}
 V(R)={V_{max}R \over \sqrt{R^2+R_c^2}},
\end{equation}

\noindent where $V_{max} = 210 \ \mathrm{km \ s}^{-1}$ is the
asymptotic maximum on the halo contribution to the rotation curve and
$R_c = 7.6 \ \mathrm{kpc}$ the core radius.

The N-body models are two-dimensional, and the gravitational potential
due to self-gravitating particles is calculated by using a logarithmic
polar grid (108 radial and 144 azimuthal cells). The N-body code we
used has been written by H. Salo \citep[for more details on the code,
  see][]{salo1991,salo2000}. The value of the gravitation softening is about
0.2 kpc on the adopted length scale. The mass of the disk
$M_{disk}=3.51 \times 10^{10} \ M_\odot$.

The disk is composed of 8 million gravitating stellar particles, whose
initial distribution is an exponential disk reaching about 10 scale
lengths. The disk and halo have nearly equal contribution to the
  rotation curve at the solar distance. The initial scale length of
the disk was about 2 kpc, but after the bar formation, it forms a twin
profile disk: the inner profile becomes steeper and the outer profile
shallower, and the exponential scale length corresponds to about 3
kpc outside the bar region. The initial value of the Toomre-parameter
$Q_T$ was 1.75.

The gas disk was modeled by inelastically colliding test particles as
was done in Paper I. The initial velocity dispersion of the gas
disk was low, about $2 \ \mathrm{km \ s}^{-1}$, but it reached
typical values in the range $5-15 \ \mathrm{km \ s}^{-1}$ during the
simulation. If collisions are omitted, the velocity dispersion of the
test particles rises much higher into the range $25-50 \ \mathrm{km
  \ s}^{-1}$. The model used in the kinematical analysis contains 40
000 gas particles initially distributed as a uniform disk with an
outer radius of 9.2 kpc.

\subsection{Evolution of the stellar component}

\begin{figure*}
\centering
\includegraphics{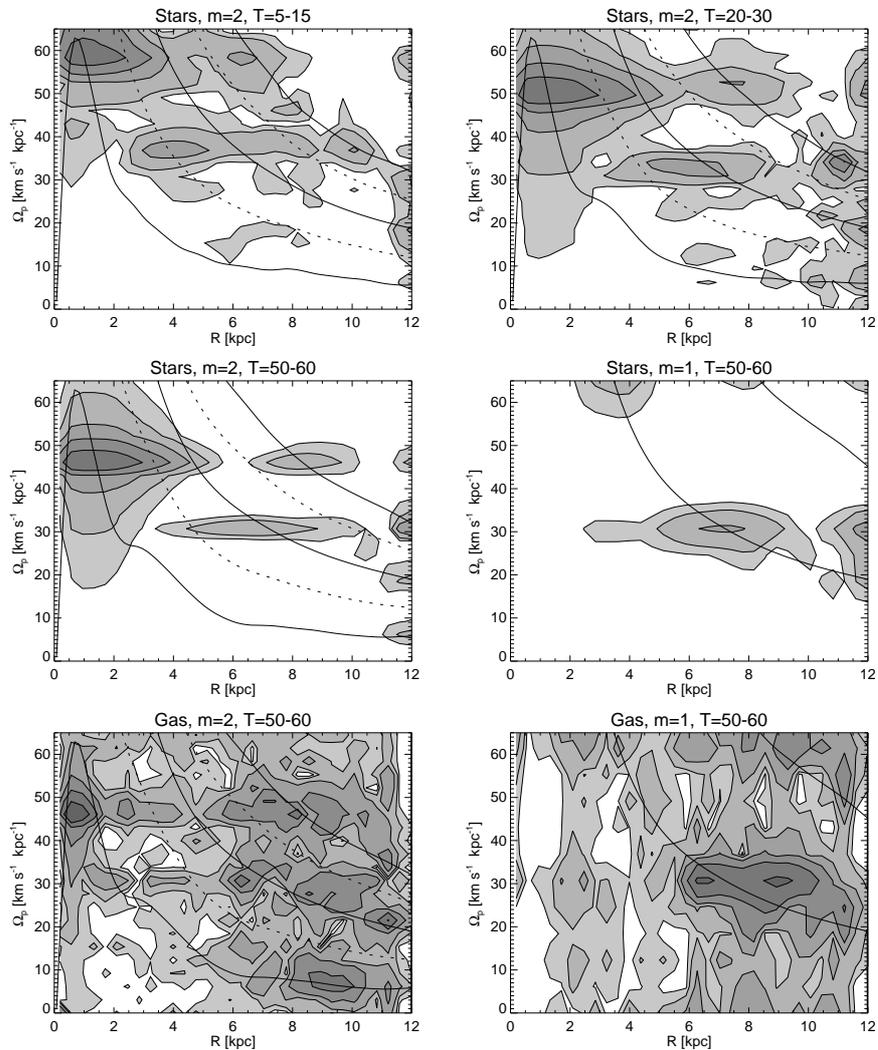}
\caption{The amplitude spectra of the relative density perturbations
  in the model disk. The frames show the amplitude spectra of the
  stellar or gas component at various times (indicated on the frame
  titles). The contour levels are 0.025,0.05,0.1,0.2,0.4, and 0.8,
  calculated with respect to the azimuthal average surface density at
  each radius. The continuous lines show the frequencies $\Omega$ and
  $\Omega \pm \kappa/m$, and the dashed curves indicate the frequencies
  $\Omega \pm \kappa/4$ in the $m=2$ amplitude spectrum.}
\label{nbody_power}
\end{figure*}

The inner regions quickly develop a small spiral (at $T \sim 2.5$),
which then evolves to a clear bar ($T \sim 5$). Its original pattern speed
$\Omega_{b}$ is about $80 \ \mathrm{km \ s}^{-1}\mathrm{kpc}^{-1}$,
meaning that when it forms it does not have an Inner Lindblad
Resonance (ILR). In its early phase the bar slows down quite quickly
($\Omega_{b} \approx 60 \ \mathrm{km \ s}^{-1}\mathrm{kpc}^{-1}$ at
$T=10$), but the deceleration rate soon settles down: $\Omega_{b}
\approx 54 \ \mathrm{km \ s}^{-1}\mathrm{kpc}^{-1} $ at $T=20$ and
$\Omega_{b} \approx 47 \ \mathrm{km \ s}^{-1}\mathrm{kpc}^{-1}$ at
$T=55$. In this model the bar's slowing down is accompanied by its growth,
and the bar can always be considered dynamically fast \citep[see
  e.g.][]{debattista2000}. Using the same method to determine the bar
length as \citet{rautiainen2008} \citep[a modification of
  one used by][]{erwin2005}, we get $R_{bar}= 4.0 \pm 0.6$ kpc at T=55
and $R_{CR}/R_{bar} = 1.2 \pm 0.2$. There is no secondary bar in this
model.

The amplitude spectra of the relative density perturbations \citep[see
  e.g.][]{masset1997,rautiainen1999} (Fig.~\ref{nbody_power}) show
that the bar mode is not the only one in the disk, but there are also
slower modes. The strongest of these modes, hereafter the S1 mode, has
an overlap of resonance radii with the bar: the corotation radius of
the bar is approximately the same as the inner 4/1-resonance radius of
the slower mode (at $T=55$ the $R_{CR}$ of the bar and the inner 4/1
resonance radius of the S1 mode are both about 4.6 kpc). This
resonance overlap does not seem to be a coincidence: when the
amplitude spectra from different time intervals are compared, one can
see that both the bar and the S1 modes slow down so that the resonance
overlap remains (see Fig.~\ref{nbody_power}). Furthermore, this
resonance overlap was the most common case in the simulations of
\citet{rautiainen1999}. Also, the S1 mode has a strong $m=1$
signal and a maximum near its corotation at 7.1 kpc. The bar mode
is also seen as a strong signal in the $m=4$ spectrum, but only inside
CR -- the spiral part seems to be almost pure $m=2$ mode. Altogether,
the signals with $m>2$ tend to be much weaker than features seen in
$m=1$ and $m=2$ amplitude spectra.

\begin{figure*}
\centering
\includegraphics{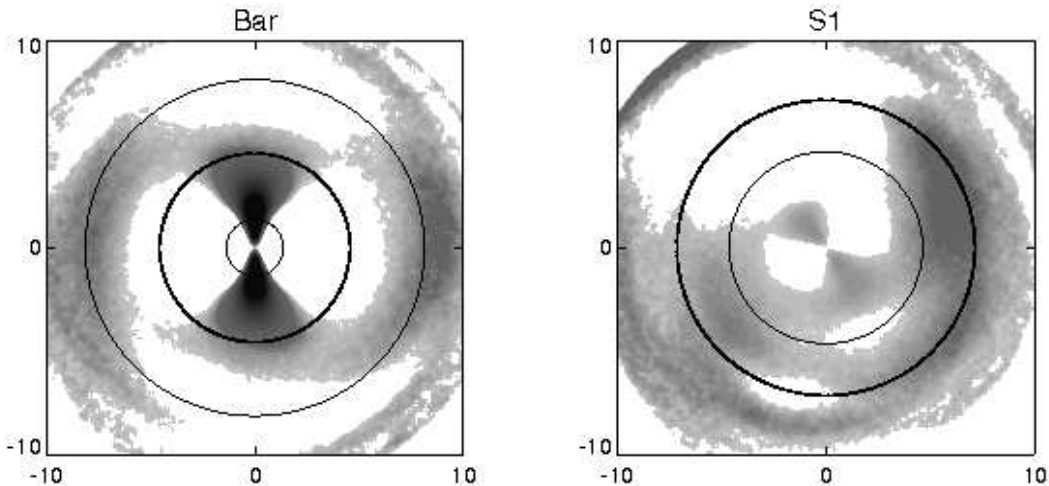}
\caption{The reconstructed modes in the stellar component (see text)
  for $T=50-60$ time interval. The enhanced density compared to the
  azimuthally averaged profile at each radius is shown.The shades of
  gray (darker corresponds to higher surface density) have been chosen
  to emphasize the features. The circles in the bar mode indicate ILR
  (1.4 kpc), CR (4.6 kpc), and OLR (8.1 kpc), whereas the inner 4/1 (4.6
  kpc) and CR (7.1 kpc) are shown for the mode S1.}
\label{mode_shapes}
\end{figure*}

We have also tried to reconstruct the shapes of the modes seen in the
amplitude spectra. This was done by averaging the surface density in
coordinate frames rotating with the same angular velocities as the
modes. No assumptions were made about the shapes of the modes. On the
other hand, one should take these reconstructions with some caution,
because the evolution of the two modes, the effect of slower (but
weaker) modes, and short-lived waves may affect them.  The results for
the bar and the S1 mode at the time interval $T=50-60$ are shown in
Fig.~\ref{mode_shapes}. The mode $\Omega_p=47 \ \mathrm{km
  \ s}^{-1}\mathrm{kpc}^{-1}$ clearly shows the bar and symmetrical
spiral structure that forms an $R_1$ outer ring or pseudoring. By the
$T=50-60$ interval, the density amplitude of the bar mode is about
15-20 per cent in the outer ring region, where the maxima and minima
have roughly the same strength. On the other hand, by $T=50-60$, the
mode $\Omega_p=31 \ \mathrm{km \ s}^{-1}\mathrm{kpc}^{-1}$ is clearly
lopsided, which is not surprising considering the signal seen in the
$m=1$ amplitude spectrum. There is a minimum with an amplitude of
about 30\% and a maximum of about 15\% at $R \approx 7$ kpc, which
corresponds to the CR of the S1 mode. Earlier, at $T \approx 20$, the
S1 mode does not have the $m=1$ characteristic but exhibits a
multiple-armed structure beyond its CR, accompanied by a clear signal
in the $m=3$ amplitude spectrum.

\subsection{The morphological changes in the gas component}

The amplitude spectra for the gas component at the interval $T=50-60$
are also shown in Fig.~\ref{nbody_power}. Due to fewer particles, they
include more noise, but otherwise they are quite similar. In addition
to the bar mode, the S1 mode is also seen, but now it is more
conspicuous in the $m=1$ spectrum.

\begin{figure*}
\centering
\includegraphics{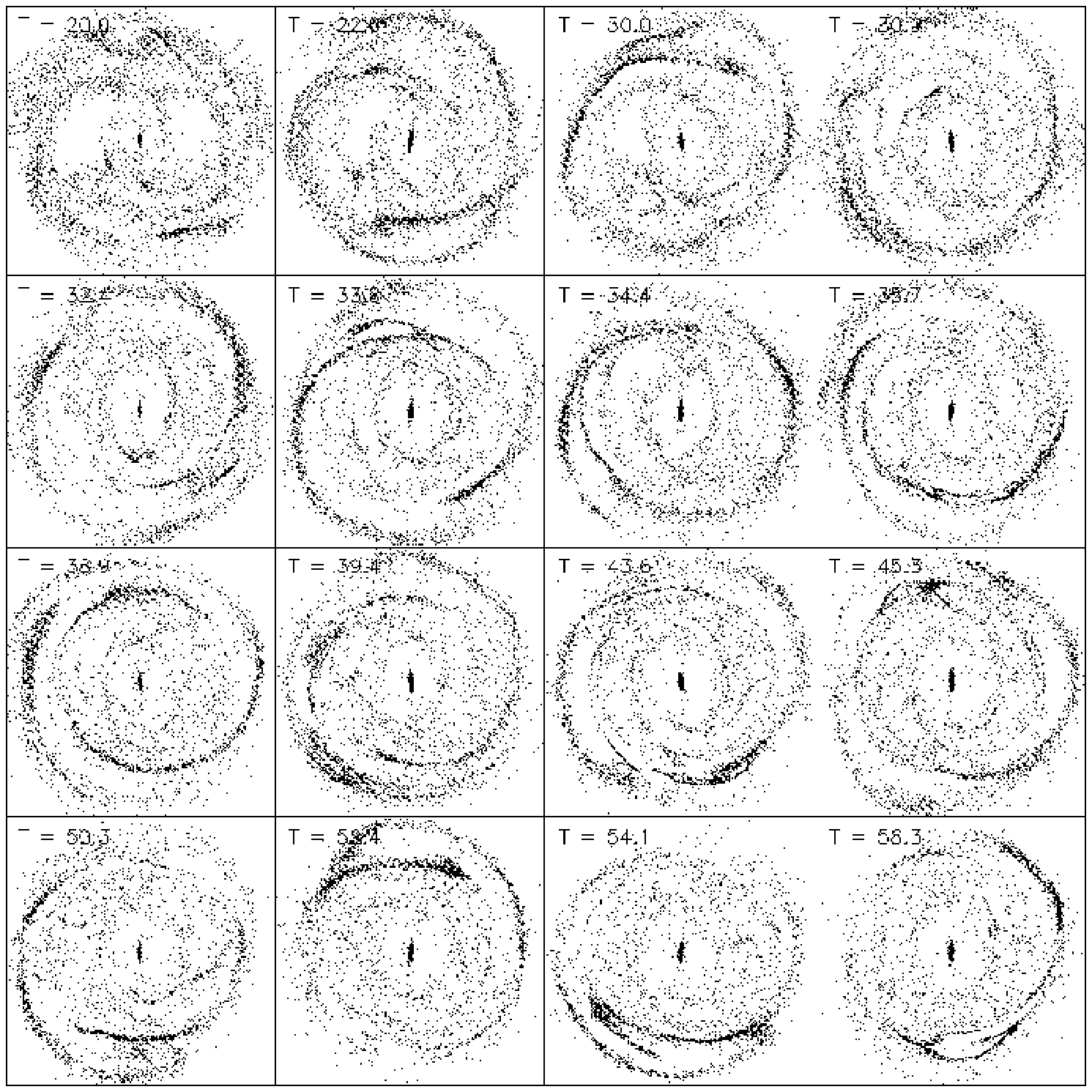}
\caption{The gas morphology at selected times. The bar is vertical in
  all frames, whose width is 20 kpc.}
\label{nbody_gasmorph}
\end{figure*}

The result of having several modes is the quite complicated
evolution of the model (see Fig.~\ref{nbody_gasmorph}): at different
times, the morphology of the outer gaseous disk can be described as
$R_1R_2'$, $R_2'$, $R_1'$ or just as open spiral arms, which can sometimes
be followed over 400 degrees. There is no evolutionary trend between
the morphological stages, since they all appear several times during the
model time span. The shape of the inner ring also changes by being
sometimes more elongated or even consisting of tightly wound pair of
spiral arms. On the broader sense, the overall Hubble stage of the
model stays the same for several Gyrs.

Although the slow modes in the stellar component can be clearly seen
outside the bar radius (about 4 kpc), they become pronounced in the
gas from $R \approx 6$ kpc.  To study their effect on the gas
morphology, we selected gas particles located at the annulus
$7<R<10$ kpc and calculated their number within every 5$^\circ$-sector
along $\theta$. Such density profiles were built for 301 moments from
the interval T=30--60 ($T \approx 3$--6 Gyr) with a step $\Delta
T=0.1$ ($\sim 10$ Myr). Earlier stages were not considered, because
then the pattern speed of the bar was changing so fast that it
complicated the analysis. At every moment the distribution of gas
density along $\theta$ was approximated by one-fold ($m=1$), two-fold
($m=2$), and four-fold ($m=4$) sinusoidal wave:

\begin{equation}
\sigma=\sigma_0+A_m\cos(m \theta +\phi_m),
  \label{sigma}
\end{equation}

\noindent where $\sigma$ is the gas density in a segment, $\sigma_0$
is the average density in the annulus, $\phi_m$ and $A_m$ are the
phase and amplitude of the corresponding sinusoidal approximation,
respectively.

\begin{figure*}
\centering
\includegraphics{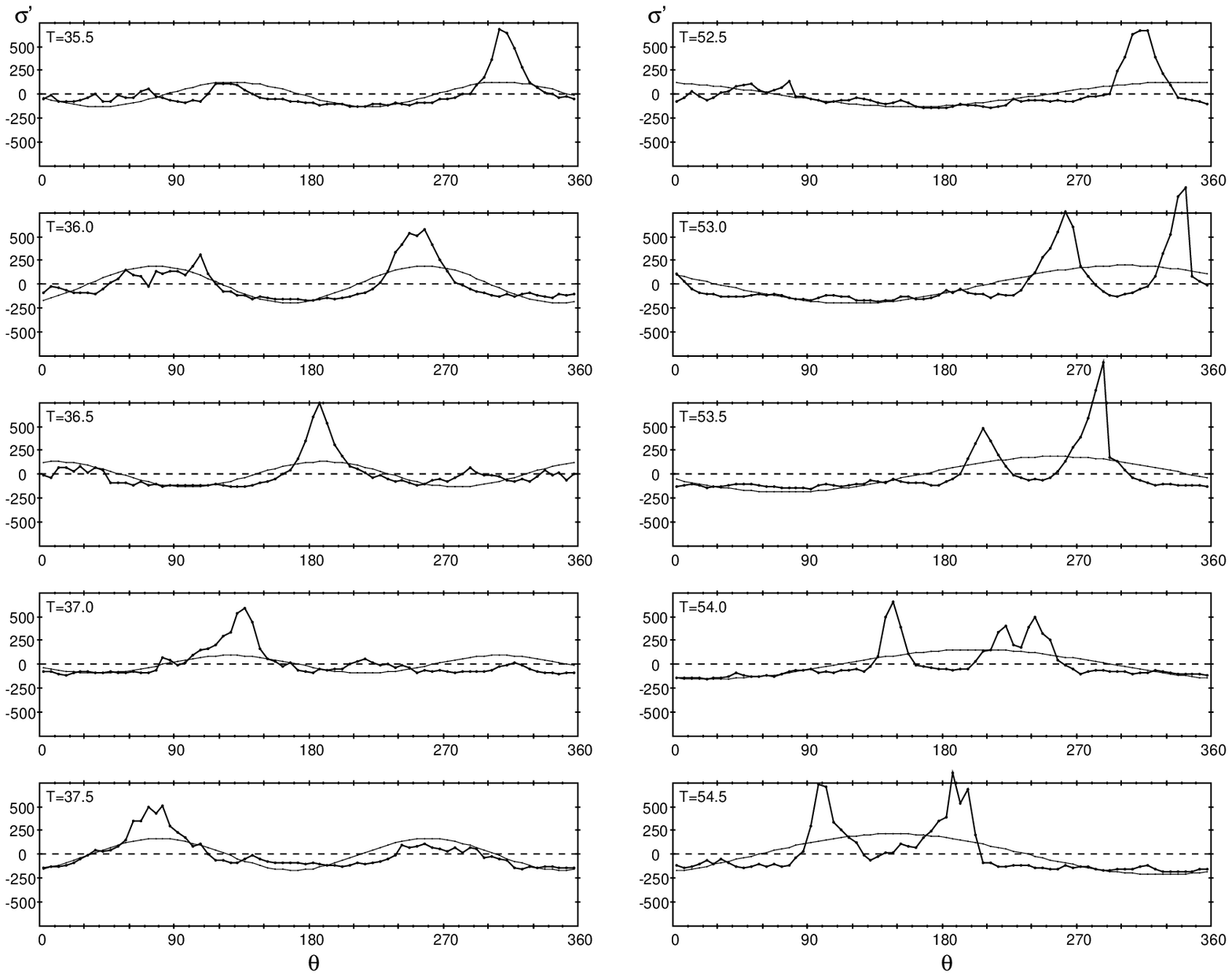}
\caption{The perturbation in the density of gas particles,
  $\sigma'=\sigma-\sigma_0$, located at the annulus $7<R<10$ kpc along
  azimuthal angle $\theta$ built for different moments. It also shows
  its approximation by two-fold (left panel) and one-fold (right
  panel) sinusoids.}
\label{humps}
\end{figure*}

Figure ~\ref{humps} demonstrates the motion of maxima in the
distribution of gas particles along $\theta$. We made the density
profiles in the reference frame co-rotating with the bar, whose major
axis is always oriented in the direction $\theta=0^\circ$.  Azimuthal
angle $\theta$ is increasing in the sense of the galactic rotation, so
the supposed position of the Sun is about $\theta=315^\circ$. To
illustrate the motion of density crests, we selected two intervals
T=35.5--37.5 and T=52.5--54.5 with a high amplitude of density
perturbation. These density profiles indicate the motion of density
maxima in the opposite direction to that of galactic rotation
(i.e. they actually rotate more slowly than the bar), which means an
increase in the phase $\phi_m$ of the sinusoidal wave
(Eq.~\ref{sigma}).

\begin{figure*}
\centering
\includegraphics{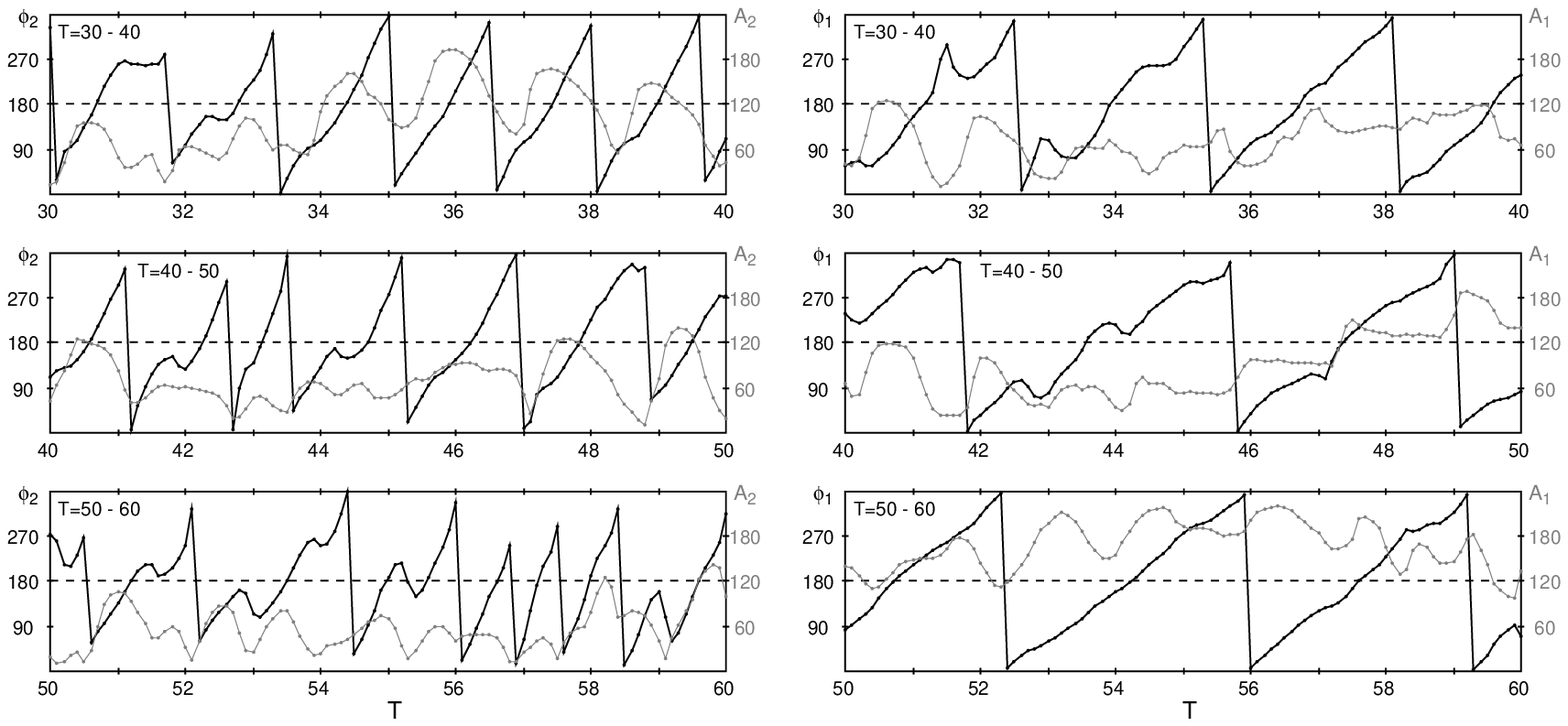}
\caption{Variations in the phase $\phi$ (black curve) and
amplitude $A$ (gray curve) of the sinusoids that approximate the
distribution of gas particles located at the distances $7<R<10$
kpc along  $\theta$. Subscripts $1$ and $2$ are  related to the
one- and two-fold sinusoids, respectively} \label{phases}
\end{figure*}

Figure ~\ref{phases} exhibits the variations in the phase $\phi_m$
and amplitude $A_m$  of the
sinusoidal wave at the time intervals T=30--40, 40--50, and 50--60.
The subscripts $1$ and $2$ are related to the one- and two-fold
sinusoids. Rotation of the density maxima causes
the sharp changes in the phase when it achieves the value of
$\phi=360^\circ$, and at the new turn its value  must fall to zero.
These changes enable us to accurately calculate the mean values
of the periods for the propagation of the sinusoidal waves, which
appear to be $P_1=3.3\pm0.4$ and $P_2=1.5\pm0.4$. Remember that
we study the density oscillations in the reference frame
co-rotating with the bar, so the period $P$ of beating
oscillations between the bar and slow modes is determined by the
relation:

\begin{equation}
 P_m=\frac{2\pi}{m(\Omega_b-\Omega_{sl})}.
  \label{P}
\end{equation}

The periods, $P_1$ and $P_2$, appear to correspond to slow modes
rotating with the pattern speeds $\Omega=28\pm2$ km s$^{-1}$ kpc$^{-1}$
and $\Omega=26\pm6$ km s$^{-1}$ kpc$^{-1}$, respectively. It is
more convenient to use simulation units here. The transformation
coefficient between them and (km s$^{-1}$ kpc$^{-1}$) is
$k=9.77$, and the value of $\Omega_b$ is $\Omega_b=4.8$ s.u. The
$m=4$ wave manifested itself in two density maxima separated by
the angle $\Delta \theta\approx90^\circ$ (Fig.~\ref{humps}, right
panel). The analysis of phase motion of four-fold sinusoid
reveals the period $P_4=0.81\pm0.15$, which corresponds to  slow
mode rotating with the speed $\Omega_{sl}=28\pm4$ km s$^{-1}$
kpc$^{-1}$ (Eq.~\ref{P}). Probably, it is mode $\Omega=28\pm4$ km
s$^{-1}$ kpc$^{-1}$ that causes the strong variations in gas
density with the periods $P_1=3.3$, $P_2=1.5$, and $P_4=0.8$ when
it works as $m=1$, $m=2$, and $m=4$ density perturbations,
respectively. This mode is well-defined in  the gas and star
power spectra made for the interval T=50-60
(Fig.~\ref{nbody_power}).

Let us have a look at the amplitude variations
(Fig.~\ref{phases}). The highest value of $A_2$
equal $A_2=200$ (particles per 5$^\circ$-sector)  is observed at
the time $T=36.0$ (left panel). On the other hand,  $A_1$
achieves its highest value of $A_1=220$  at the time $T=56.5$
(right panel).  Amplitude $A_4$ reaches its maximum value of
$A_4=180$  at the time interval $T=53-55$. Thus, the highest
values of the amplitudes $A_1$, $A_2$, and $A_4$ are nearly the
same.

Figure ~\ref{humps} (left panel)  indicates the growth of the
amplitude  of m=2 perturbation under a specific orientation of the
density clumps.  The amplitude  of the sinusoidal wave is
at its maximum at the moments $T=36.0$ and 37.5 when the density clumps
are located near the bar's minor axis, $\theta=90^\circ$ and
270$^\circ$.  This growth is also seen in Fig.~\ref{phases} (left
panel) for the interval T=30-40: the amplitude $A_2$ is at its maximum
at the moments when $\phi_2\approx 180^\circ$. This phase
corresponds to the location of  maxima of $m=2$ sinusoid at
$\theta=90^\circ$ and $\theta=270^\circ$ (Eq.~\ref{sigma}).

Our analysis revealed slight variations in the speed of the strongest
slow mode, and they depend on its orientation with respect to the bar:
Fig.~\ref{phases} (left panel) shows that the tilt of the phase
curve, $\phi_2(t)$, is variable. We can see that the slow mode rotates
a bit faster when $\phi_2\approx 180^\circ$ (density clumps are near
the bar's minor axis) and more slowly when $\phi_2=0$ or 360$^\circ$ (the
clumps are near the bar's major axis). Probably, the variations in the speed
of the slow mode are connected with the change in the form of the
density crests due to tidal interaction between the bar mode
(bar+$R_1$-ring) and the slow mode.


\section{Kinematics of gas particles. Comparison with observations}

\subsection{Momentary and average velocities}

We start our kinematical study with the interval T=50--60 (5--6 Gyr in
physical time). At this period the bar rotates with a nearly constant
pattern speed of $\Omega_b=47$ km s$^{-1}$ kpc$^{-1}$ which simplifies
the analysis. The interval T=50--60 also provides the best
agreement between the model and observed velocities.

We determined the positions and velocities of gas particles at
101 moments separated by the step $\Delta T=0.1$.  For each
moment we selected gas particles located within the boundaries of
the stellar gas complexes and calculated their mean velocities
and velocity dispersions. To determine the positions of the complexes,
we need to choose the position angle of the Sun with respect to
the bar elongation, $\theta_b$. In this section we adopted the
value of $\theta_b=45^\circ$, which gives the best fit between the
model and observed velocities.

\begin{figure*}
\centering
\includegraphics{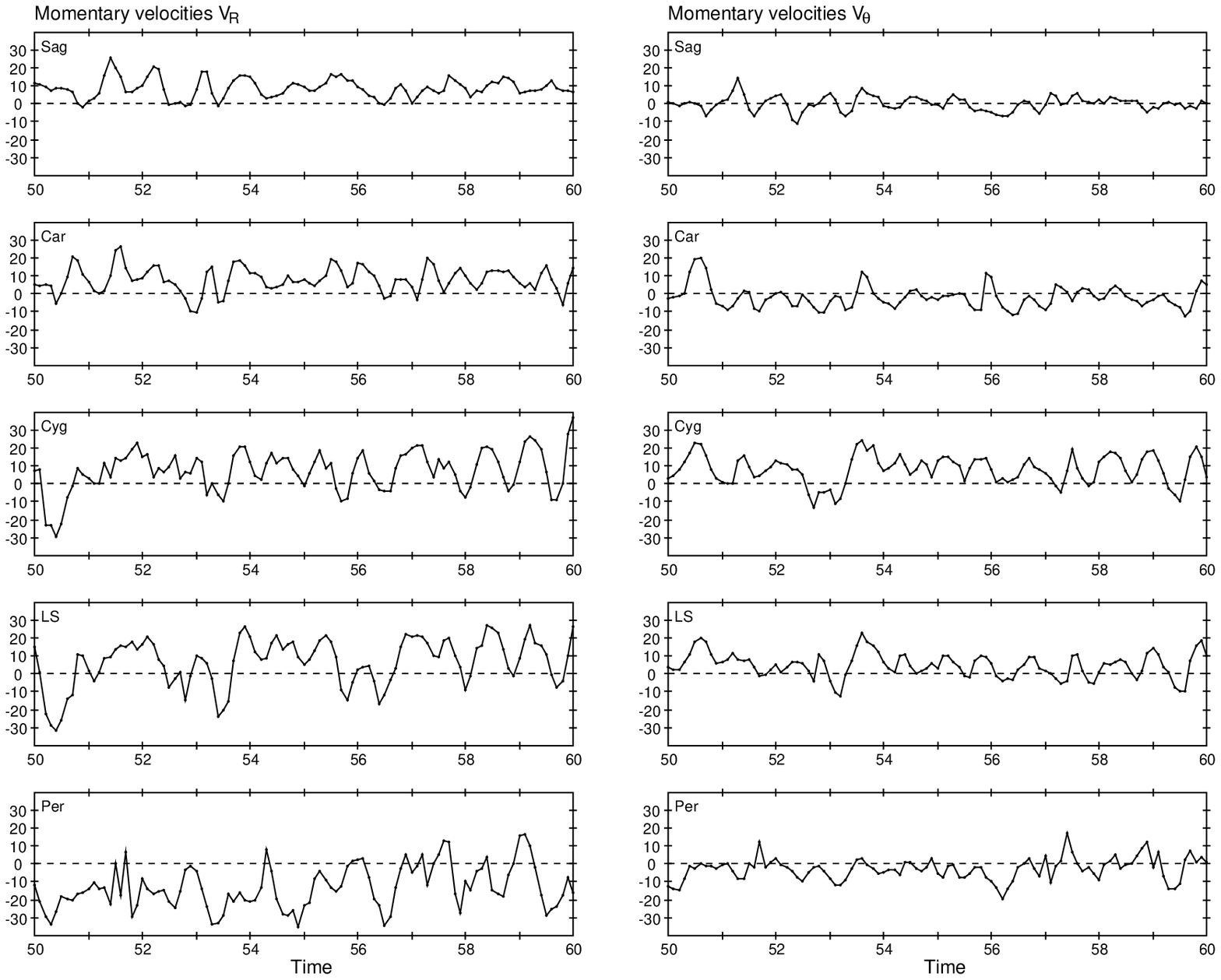}
\caption{Variations in the mean velocities of gas particles
located within the boundaries of the stellar-gas complexes. The
left panel is related to the radial component $V_R$ and the
right one to the azimuthal one $V_\theta$.} \label{vel-var}
\end{figure*}

Figure ~\ref{vel-var} shows the variations in the mean residual
velocities, $V_R$ and $V_\theta$, calculated for five complexes at
different moments. The residual velocities were computed as
differences between the model velocities and the velocities due to the
rotation curve.  It is clearly seen that the momentary velocities
oscillate near the average values within the limits of $\sim\pm20$ km
s$^{-1}$. Two processes are probably responsible for these
oscillations. The first is the slow modes that cause a quasi-periodic
low in the velocity variations.  The second process is likely
connected with the short-lived perturbations, e.g. from transient
spiral waves in the stellar component. The averaging of velocities
over long time interval reduces the influence of slow modes and
occasional perturbations.

Table~\ref{table-average1} represents the average values of the
momentary residual velocities, $\overline{V_R}$ and
$\overline{V_\theta}$, calculated over 101 moments. It also gives the
average values of velocity dispersions, $\overline{\sigma_R}$ and
$\overline{\sigma_\theta}$, and the average number of particles
$\overline{n}$ in the complexes. Since the bar has two tips, we
calculated velocities for two opposite positional angles,
$\theta_b=45^\circ$ and $\theta_b=225^\circ$, and used their mean
values. The averaged residual velocities are determined with the
errors of 0.4--1.4 km s$^{-1}$. The relatively low level of the errors
is due to the large number of moments considered (N=101).

\begin{table}
\centering
 \caption{Model residual velocities  averaged on interval T=50-60}
  \begin{tabular}{lccccc}
  \hline
   Region  &  $\overline{V_R}$ & $\overline{\sigma_R}$ & $\overline{V_\theta}$ &$\overline{\sigma_\theta}$ &  $\overline{n}$\\
             & km s$^{-1}$  &  km s$^{-1}$ &  km s$^{-1}$&  km s$^{-1}$ &   \\
 \hline
Sagittarius &  8.5& 7.2& 0.1&5.9& 70\\
Carina      &  7.5& 7.6&-2.0&6.6&158\\
Cygnus      &  6.8&10.1& 8.2&6.5&108\\
Local System&  6.8&11.7& 4.8&6.7&112\\
Perseus     &-12.5&11.9&-2.9&6.5& 70\\
\hline
\end{tabular}
\label{table-average1}
\end{table}

When comparing Tables ~\ref{table-average1} and
~\ref{observations}, one can see that our model reproduces the
directions of the radial and azimuthal components of the residual
velocities in the Perseus and Sagittarius regions and in the Local
System. We succeed in the Sagittarius region where our model
reproduces the observed velocities with the accuracy 1.4 km s$^{-1}$.
Unfortunately, in the Perseus region the model residual velocity
$|\overline{V_R}|$ is too high, and the difference between the
model and observed velocities achieves 5.8 km s$^{-1}$ there. Our
model can also reproduce the positive $\overline{V_R}$ velocity in the
Local System, which deviates only 1.5 km s$^{-1}$ from the observed
one.

We now consider  the mean  difference between the model and
observed velocities $\Delta V$ calculated for the radial and
azimuthal components:

\begin{equation}
\Delta V^2=\frac{\sum^k_1 \left\{ (\overline{V_R}-V_{R\mbox{
obs}})^2+ (\overline{V_\theta}-V_{\theta\mbox{ obs}})^2 \right\}
 }{2k},
\label{delta_v}
\end{equation}

\noindent where $k$ is a number of complexes. The value of
$\Delta V$ computed for  three complexes (the Sagittarius and
Perseus regions and the Local System) equals $\Delta V=3.3$ km
s$^{-1}$. Another situation is observed in the Carina and Cygnus
regions where we cannot even reproduce the direction of the
observed residual velocities. The mean value of the velocity
deviations achieves $\Delta V=13.3$ km s$^{-1}$ there.

\begin{figure*}
\centering
\includegraphics{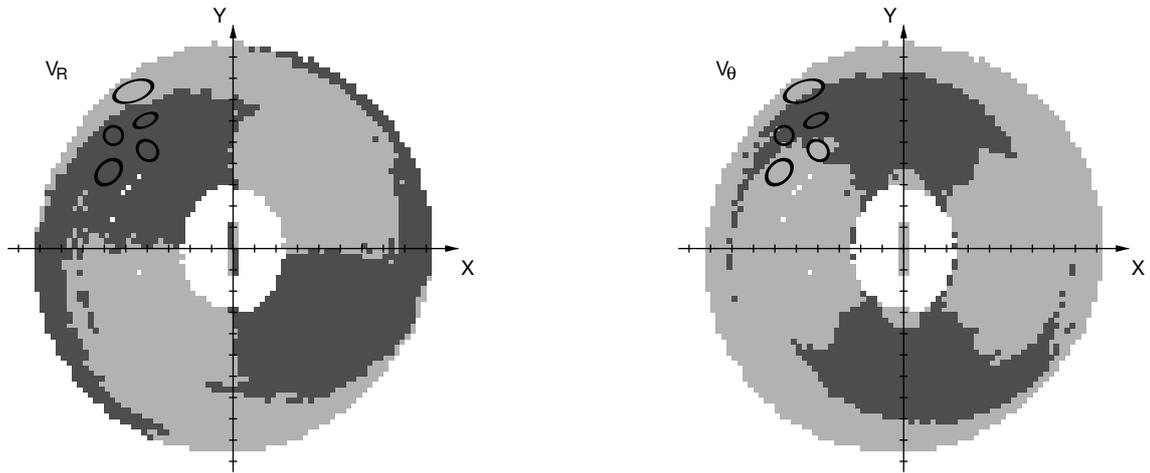}
\caption{Distribution of the negative and positive average
residual velocities calculated in squares. The squares with
positive velocities are shown in black, while those with negative
ones are given in gray. Only squares that satisfy the condition
$n>\overline{m}/2$ are depicted. The left panel represents the
radial velocities, while the right one shows the azimuthal ones. It
also demonstrates the boundaries of the stellar-gas complexes.
The position angle of the Sun is supposed to be
$\theta_b=45^\circ$. The bar is oriented along the Y-axis, the
galaxy rotates clockwise, and a division on the $X$- and $Y$-axis
corresponds to 1 kpc.}  \label{vel-average}
\end{figure*}

To demonstrate the distribution of the average velocities on the
galactic plane, we divided the area ($-10<x<+10$, $-10<y<+10$
kpc) into small  squares of a size $0.250\times0.250$ kpc. For
each square we calculated the average values of the residual
velocities of gas particles. Then we averaged residual velocities
over 101 moments for the interval T=50--60. The average residual
velocities in squares are shown in Fig.~\ref{vel-average}. We
depicted only squares that contain high enough number of particles,
$n>\overline{m}/2$, where $n$ is the number of particles
accumulated in a square over 101 moments but $\overline{m}$ is
their number averaged over all squares, $\overline{m}=463$.

In Paper I we have built similar figures for models with
analytical bars. Two different moments were considered:
when the broken rings (pseudorings) were observed and when they
transformed into pure rings. The pseudorings and pure rings
created different kinematical pictures. We connected the main
kinematical features of the pseudorings with the gas outflow and those
of the pure rings with the resonance. The distribution of the negative
and positive velocities obtained for N-body simulations
(Fig.~\ref{vel-average}) strongly resembles that of the pseudorings in
models with analytical bars, giving support to the ``averaging
process'' adopted here. This similarity suggests there is
gas outflow in the present model (see also Sect. 4.4).

\subsection{Velocities in the complexes under different values of the solar position angle $\theta_b$}

\begin{figure}
\resizebox{\hsize}{!}{\includegraphics{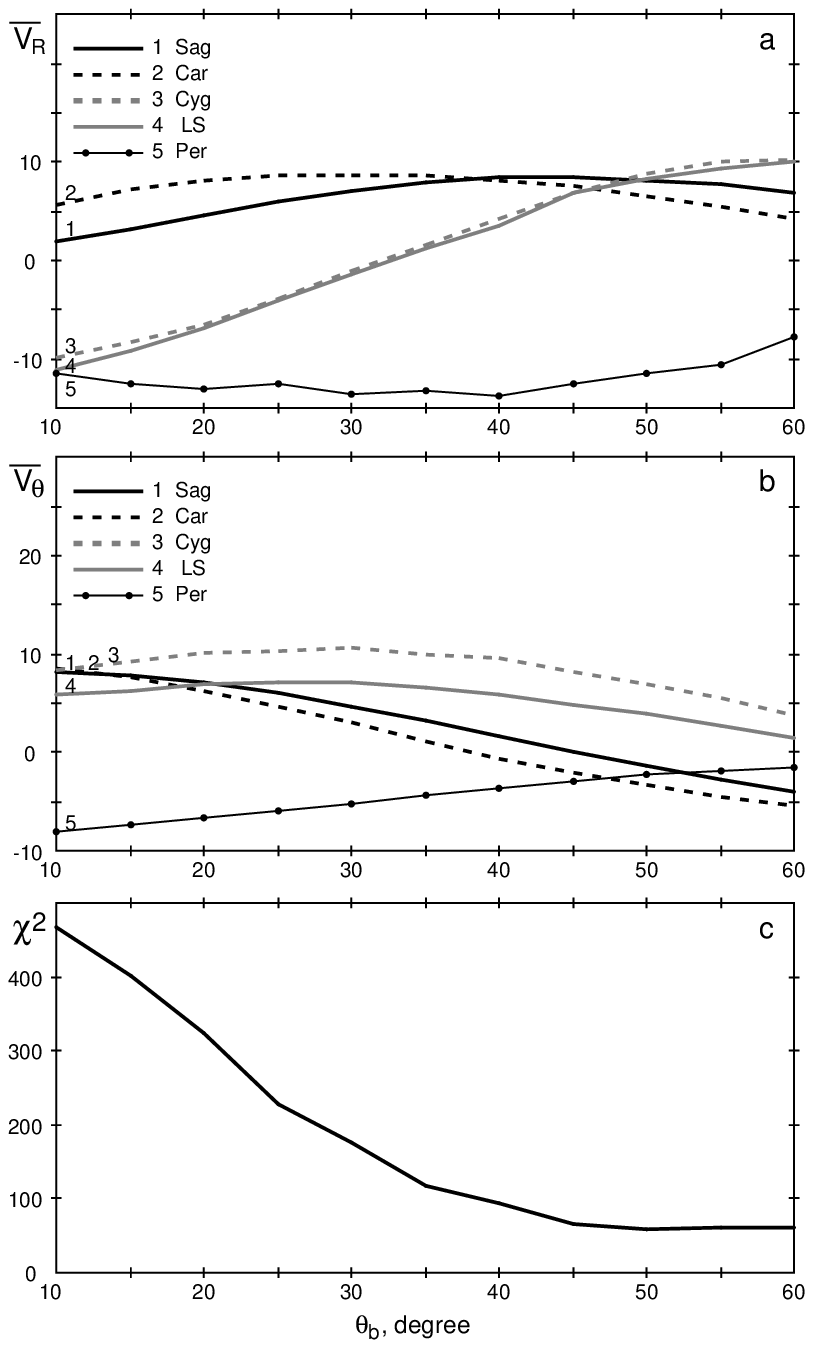}}
\caption{Dependence of the average residual  velocities,
$\overline{V_R}$ (a) and  $\overline{V_\theta}$ (b), and the
$\chi^2$-function (c) on the solar position angle $\theta_b$. The
curves calculated for different complexes are shown by different
lines. The $\chi^2$-function was computed for three complexes:
the Perseus and Sagittarius regions and the Local System.}
\label{v-theta_b}
\end{figure}

We studied the dependence of the average residual velocities
$\overline{V_R}$ and $\overline{V_\theta}$ on the solar position angle
$\theta_b$. Figure ~\ref{v-theta_b}ab shows 5 curves that demonstrate
the velocity changes in 5 complexes. The sharpest changes in the
radial velocity $\overline{V_R}$ are observed in the Local System and
in the Cygnus region, and the radial velocities in the other complexes
depend only weakly on the choice of $\theta_b$.  As for the azimuthal
component, the strongest changes can be seen in the Sagittarius,
Carina, and Perseus regions, but the velocity changes are modest in
other complexes.  Practically speaking, the optimal value of
$\theta_b$ provided the best agreement between the model and observed
velocities is determined by the radial velocity in the Local System
and by the azimuthal velocity in the Sagittarius region. These
velocities achieve their observed values of $V_R=5.3$ and
$V_\theta=-1$ km s$^{-1}$ under $\theta_b=43^\circ$ and
$\theta_b=48^\circ$, respectively.

We now consider the sum of square differences between the model and
observed velocities, $\chi^2$, obtained for the radial and azimuthal
components under different values of $\theta_b$.  Figure
~\ref{v-theta_b}c shows the $\chi^2$-function computed for three
complexes: the Perseus and Sagittarius regions and the Local
System. It is clearly seen that $\chi^2$ achieves its minimum values
at the interval $\theta_b >40^\circ$. We chose $\theta_b=45^\circ$ as
the optimal value because it reproduces the observational velocity
$V_\theta=-1$ km s$^{-1}$ well in the Sagittarius region. Models with
analytical bars in Paper I gave the same result.

\subsection{Analysis of periodicity in oscillations of the momentary velocities}

Now we approximate the oscillations in the radial and azimuthal
components of the momentary velocities, $V_R$ and $V_\theta$
(Fig.~\ref{vel-var}), by the sinusoidal law:

\begin{equation}
V_R (\textrm{or }V_\theta)=A_1\sin(2\pi T/P)+A_2\cos(2\pi T/P),
\end{equation}

\noindent where $P$ is a period of oscillations,
$A_0=\sqrt{A_1^2+A_2^2}$ is an amplitude of oscillations, and $T$ is
time counted from $T_0=50$.

\begin{figure}
\resizebox{\hsize}{!}{\includegraphics{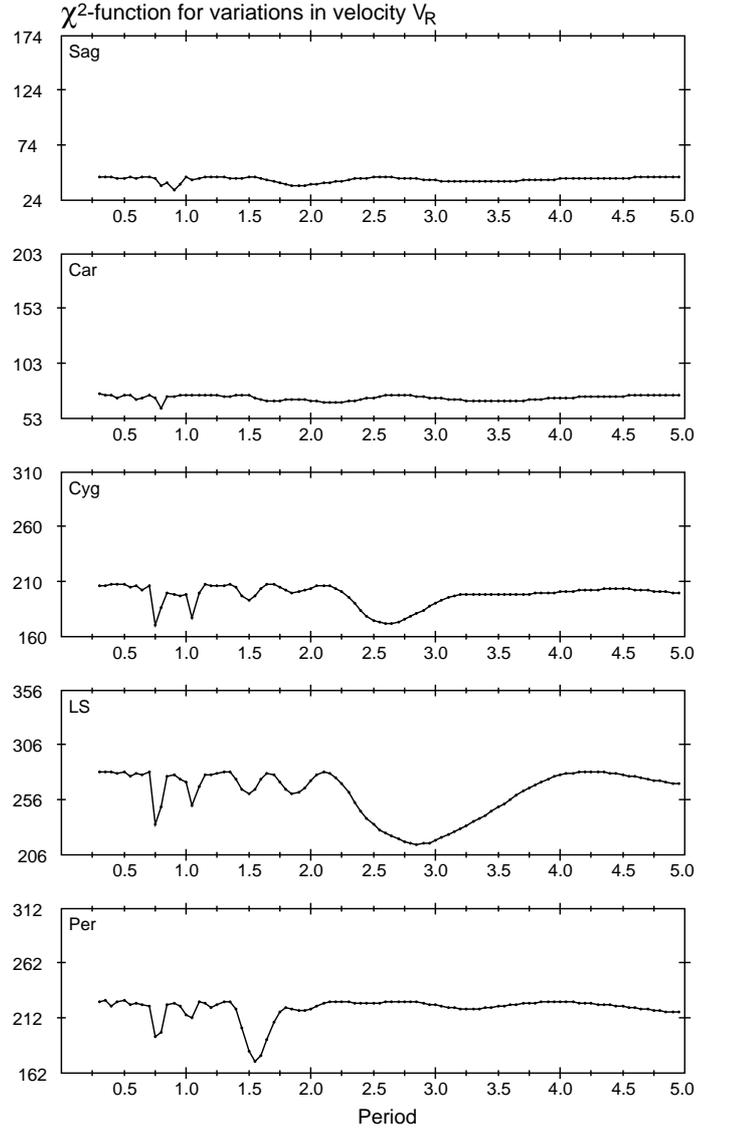}}
\caption{$\chi^2$-functions built for studying periodicity in
the oscillations of the radial velocities, $V_R$, in 5 complexes.
The minima on the curves must correspond to the best periods in
approximation of the velocity oscillations.} \label{period}
\end{figure}

We use the standard least square method to solve the system of 101
equations, which are linear in the parameters $A_1$ and $A_2$ for each
value of nonlinear parameter $P$. We then determine the value of $P$
that minimizes the sum of squared normalized residual velocities
$\chi^2$. Figure ~\ref{period} presents the $\chi^2$-curves built for
the oscillations of the radial velocity in 5 complexes, but the curves
made for the azimuthal velocities have no conspicuous minima. It is
clearly seen that $\chi^2$-curves demonstrate deep minima in the
Cygnus and Perseus regions and in the Local System. These minima
correspond to the best periods in approximating the velocity
oscillations that have the following values: $P=2.7\pm0.4$ in the
Cygnus region, $P=2.9\pm1.0$ in the the Local System, and
$P=1.6\pm0.2$ in the Perseus region. We have already obtained period
$P=1.5$ when studying density oscillations on the galactic periphery
(Sect. 3.3). Probably, the strongest slow mode $\Omega=28$ km s$^{-1}$
kpc$^{-1}$ is also responsible for the velocity oscillations: the
beating oscillations between the bar mode and a two-armed pattern
rotating with the speed $\Omega=28$ km s$^{-1}$ kpc$^{-1}$ must have
the period of $P=1.6$ and those calculated for one-armed perturbation
have a period of $P=3.2$. (Eq.~\ref{P}). Some of the small differences
between the pattern speeds derived from the amplitude spectra and
those obtained from kinematical analysis may be due to tidal
interaction in the stellar component between the bar and slow modes.

\subsection{Evolutional aspects of kinematics at the time interval
T=30-60}

Let us compare the average residual velocities calculated for
different time intervals T=30--40, 40--50, and 50--60
(Tables~\ref{table-average1}, \ref{table-average2}, and
\ref{table-average3}). Generally, most changes in the residual
velocities do not exceed 4.0 km s$^{-1}$ and are likely caused by
occasional perturbations. On the other hand, radial velocities
$\overline{V_R}$ in the Local System and in the Cygnus region
demonstrate the ongoing growth, which can be connected with the
evolution of the outer rings.

\begin{figure*}
\resizebox{\hsize}{!}{\includegraphics{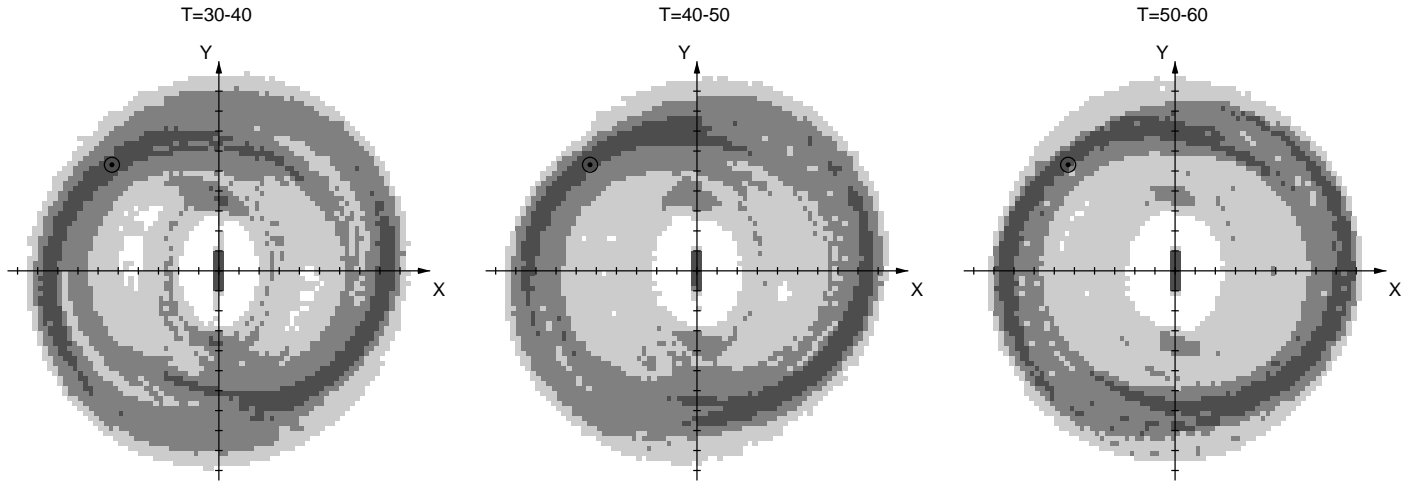}}
\caption{The surface density of gas particles averaged in squares
at the time intervals T=30--40, 40--50, and 50--60.  The
light-gray, dark-gray, and  black colors represent squares
containing the increasing number of particles:
$n>\overline{m}/2$, $n>\overline{m}$, and $n>2\overline{m}$,
respectively. The position of the Sun is shown by the specific
symbol. The bar is oriented along the Y-axis, the galaxy rotates
clockwise, and a division on the $X$- and $Y$-axis corresponds to 1
kpc.} \label{density_30-60}
\end{figure*}

Figure ~\ref{density_30-60} shows the surface density of gas particles
averaged in squares at different time intervals. The average density
was calculated in the reference frame that rotates with the speed of
the bar. The light-gray, dark-gray, and black colors represent squares
containing the increasing number of particles, $n>\overline{m}/2$,
$n>\overline{m}$, and $n>2\overline{m}$, respectively, where $n$ is
the number of particles accumulated in a square over 101 moments and
$\overline{m}$ is their number averaged over all squares,
$\overline{m}=463$. It is clearly seen that the major axis of the
outer ring $R_2$ changes its orientation: it goes $\alpha\sim20^\circ$
ahead of the bar at the interval T=30--40, but this angle increases to
$\alpha\sim45^\circ$ at the intervals T=40--50 and T=50--60. Moreover,
the outer ring changes its morphology: we can identify two outer rings
of classes $R_1$ and $R_2$ at the interval T=30--40, while there is
only one outer ring with an intermediate orientation of
$\alpha\approx45^\circ$ at the intervals T=40--50 and 50--60. Its
shape becomes rounder at the interval T=50--60.

Let us consider more thoroughly the distribution of gas particles at
the interval T=50--60 (Fig.~\ref{density_30-60}). It is clearly seen
that the surface density of gas particles at the distance range of
$R=6-9$ kpc is nearly twice the average density all over the disk
$\overline{m}$. The density perturbation inside the outer ring can be
approximated by two spiral arms with a pitch angle of
$i=6\pm1^\circ$. The  density perturbation inside them reaches to
100 per cent with respect to the average gas density in the disk.
This is considerably larger than the density
perturbation seen in the stellar component (15-20 per cent).

\begin{figure}
\resizebox{\hsize}{!}{\includegraphics{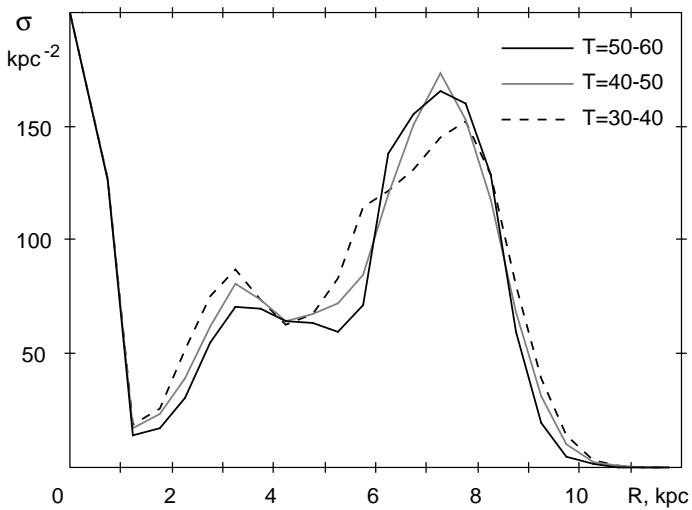}}
\caption{Profiles of the surface density of gas particles
averaged at the intervals T=30--40, 40--50, and 50--60.}
\label{density-profile}
\end{figure}

Figure ~\ref{density-profile} shows the profiles of the surface
density of gas particles averaged at the different time
intervals. We can see the growth of the density hump at the
distance of $R\approx7$ kpc, which indicates the growth of the
outer ring. In contrast, the hump at $R\approx 3$ kpc is
decreasing, which reflects  the weakening of the inner ring. At
the interval $T=50$--60, the maximum in  the gas density
distribution is located at the distance $R=7.3$ kpc, which is just
in the middle between the outer 4/1 resonance (6.4 kpc) and the
OLR (8.1 kpc) of the bar.

Tables~\ref{table-average1}, \ref{table-average2},
\ref{table-average3} also represent  the  velocity dispersions of
gas particles in the stellar gas complexes. We can see that their
average values stay at nearly the same level of
$\overline{\sigma_R}=9.7\pm0.1$ and
$\overline{\sigma_\theta}=6.3\pm0.2$ km s$^{-1}$ during the period
T=30--60. The maximum growth, which does not exceed $\sim 20$ per
cent, is observed in the Perseus region. The model velocity
dispersions somewhat exceed the observed values derived for
OB-associations in the stellar-gas complexes, $\sigma_{R\mbox{
obs}}=7.7$ and $\sigma_{\theta\mbox{ obs}}=5.2$ km s$^{-1}$, but
this difference is below 30 per cent.

\begin{table}
\centering
 \caption{Model residual velocities  averaged on interval T=30-40}
  \begin{tabular}{lccccc}
  \hline
   Region  &  $\overline{V_R}$ & $\overline{\sigma_R}$ & $\overline{V_\theta}$ &$\overline{\sigma_\theta}$ &  $\overline{n}$\\
             & km s$^{-1}$  &  km s$^{-1}$ &  km s$^{-1}$&  km s$^{-1}$ &   \\
 \hline
Sagittarius & 10.2& 8.4& 1.8&6.9& 84\\
Carina      & 10.4& 8.7& 1.2&7.8&184\\
Cygnus      & -0.7& 9.9&10.8&5.4& 87\\
Local System& -1.0&11.6& 7.7&5.9& 90\\
Perseus     &-11.3& 9.7&-2.5&5.6& 83\\
\hline
\end{tabular}
\label{table-average2}
\end{table}
\begin{table}
\centering
 \caption{Model residual velocities  averaged on interval T=40-50}
  \begin{tabular}{lccccc}
  \hline
   Region  &  $\overline{V_R}$ & $\overline{\sigma_R}$ & $\overline{V_\theta}$ &$\overline{\sigma_\theta}$ &  $\overline{n}$\\
             & km s$^{-1}$  &  km s$^{-1}$ &  km s$^{-1}$&  km s$^{-1}$ &   \\
 \hline
Sagittarius &  8.9& 7.4& 0.8&5.7& 69\\
Carina      &  9.1& 7.4&-1.0&6.8&164\\
Cygnus      &  3.9&10.4&11.4&5.7&126\\
Local System&  5.6&12.5& 8.3&6.7&121\\
Perseus     &-15.2&11.1&-1.9&5.5& 50\\
\hline
\end{tabular}
\label{table-average3}
\end{table}

\section{Conclusions}

We have presented N-body simulations that reproduce the kinematics of
OB-associations in the Perseus and Sagittarius regions and in the
Local System. The velocities of gas particles averaged over large time
intervals (1 Gyr or 8 bar rotation periods) reproduce the directions
of the observed velocities in these regions. The mean difference
between the model and observed velocities calculated for the radial
and azimuthal components is $\Delta V=3.3$ km s$^{-1}$ there.

The galactic disk in our model includes two subsystems. The behavior
of the stellar subsystem is modeled by 8 million gravitating
collisionless particles. The stellar disk quickly forms a bar.  Its
original pattern speed is quite high, but it first quickly decreases
and then moves to a slow decrease with $\Omega \approx 50 \ \mathrm{km
  \ s}^{-1} \mathrm{kpc}^{-1}$ for several Gyrs. With our favored
value of the solar distance, $R_0=7.5 \ \mathrm{kpc}$, this sets us
close to the OLR ($R_{OLR}=8.1 \ \mathrm{kpc}$). This agrees with
studies of local stellar velocity distribution
\citep{dehnen2000,fux2001,minchev2009}, although they tend to set the
OLR slightly inside $R_0$. The optimal value of the solar position
angle $\theta_b$ providing the best agreement between the model and
observed velocities is $\theta_b=45\pm5^\circ$. The bar is quite long
($R_{bar} \approx 4.0$ kpc), but both its size and orientation are
consistent with the parameters derived from infrared observations
\citep{benjamin2005,cabrera-lavers2007}.

The stellar disk also creates an outer ring of class $R_1$ rotating
with the pattern speed of the bar, and the corresponding density
perturbation amounts to 15--20 per cent of the average density at the
same distance. Besides the bar, the stellar disk includes several slow
modes. The strongest of these rotates with the pattern speed of
$\Omega \approx 30 \ \mathrm{km \ s}^{-1} \mathrm{kpc}^{-1}$ and is
often clearly lopsided.

The gas subsystem is modeled by 40 000 massless particles that move in
the potential created by the stellar particles (and analytical bulge
and halo) and can collide with each other inelastically.  The gas disk
forms an outer ring that exhibits quasi-periodic changes in its
morphology because it has several modes. One can identify elements of
$R_1$- and $R_2$-morphology, and the outer ring can often be
classified as $R_1R_2'$. The gas density perturbation inside the ring
can be approximated by two spiral arms with the pitch angle of
$i=6\pm1^\circ$.

The models with analytical bars (Paper I) reproduced the residual
velocities well in the Perseus and Sagittarius regions.  We explained this
success by the resonance between the relative orbital rotation of the
bar and the epicyclic motion. The Sagittarius region must be located
slightly inside the OLR where resonance orbits are elongated
perpendicular to the bar, whereas the Perseus region must lie outside
the OLR where periodic orbits are oriented along the bar. However,
models with the analytical bar failed dramatically with the Local System
where they yielded only negative radial velocities $-24<V_R<-16
\ \mathrm{km \ s}^{-1}$, whereas the observed value is $+5.3
\ \mathrm{km \ s}^{-1}$. The success of N-body simulations with the Local
System is likely due to the gravity of the stellar $R_1$-ring, which
is omitted in models with analytical bars.

To study the effects of the gravity of the $R_1$-ring  we
construct more simple models with a ``time averaged bar
potential''. This was done by calculating the average density
distribution  in the frame rotating with the bar. This process
should average out most of the effect of slower modes and leave
bar and the $R_1$-ring that corotates with the bar. The
preliminary study shows that momentary velocities in these models
are in a good agreement with the average velocities in the
present N-body simulation. The detailed description of these models
will be done in our next paper.

To simplify the analysis at this point we are forced to ignore a lot
of processes which are important at such long time interval as 6
Gyr. We do not consider the accumulation of gas at the galactic
center, the transitions between the gas and stellar subsystems,
resonant interaction between the bar and halo, or the minor mergers
and satellite accretion. Considering the effects of these processes
may be done in a later phase.

\begin{acknowledgements}

We want to
thank H. Salo who wrote the simulation code we have used in this
study. This work was partly supported by the Russian Foundation for Basic
Research (project nos.~10\mbox{-}02\mbox{-}00489). 

\end{acknowledgements}


\begin{thebibliography}{}

\bibitem[Athanassoula et al.(2009)]  {athanassoula2009}
  Athanassoula E., Romero-G{\'o}mez M., Masdemont
  J.~J.,\ 2009, MNRAS, 394, 67
\bibitem[Bagley et al.(2009)]{bagley2009} Bagley, M., Minchev, I., 
\& Quillen, A.~C.\ 2009, \mnras, 395, 537 
\bibitem[Barbier-Brossat and Figon(2000)]{barbierbrossat2000}
  Barbier-Brossat M., Figon P.,\ 2000, A\&AS, 142, 217
\bibitem[Benjamin at al.(2005)] {benjamin2005} Benjamin R.~A.,
Churchwell E., Babler B.~L. et al.\ 2005, \apj, 630, L149.
\bibitem[Berdnikov et al.(2000)]{berdnikov2000}
  Berdnikov L.~N., Dambis A.~K., Vozyakova O.~V., \ 2000, A\&AS, 143, 211
\bibitem[Bissantz
\& Gerhard(2002)]{bissantz2002} Bissantz, N., \& Gerhard, O.\ 2002, \mnras, 330, 591
\bibitem[Blaha \& Humphreys(1989)]{BlahaHumphreys1989}
  Blaha, C., \&  Humphreys, R.~M. 1989, \aj, 98, 1598
\bibitem[Blitz \& Spergel(1991)]{blitz1991} Blitz, L., \& Spergel, D.~N.\ 1991, ApJ, 379, 631
\bibitem[Blitz et al.(1993)]{blitz1993}
  Blitz L., Binney J., Lo K.~J., Bally J., Ho P.~T.~P., \ 1993, Nature, 361, 417
 \bibitem[Bobylev at al.(2007)]{bobylev2007}
 Bobylev V.~V., Bajkova A.~T., Lebedeva S.V.,\ 2007, Astron. Lett., 33, 720
\bibitem[Burton \& Gordon(1978)]{burton1978}
  Burton W.~B., Gordon M.~A., \ 1978, A\&A, 63, 7
\bibitem[Buta(1986)]{buta1986}
  Buta R.,\ 1986, ApJS, 61, 609
\bibitem[Buta(1995)]{buta1995}
  Buta  R.,\ 1995, ApJS, 96, 39
\bibitem[Buta \& Combes(1996)]{buta1996}
  Buta R., Combes F., \ 1996, Fund. Cosmic Physics, 17, 95
\bibitem[Buta et al.(2007)]{buta2007}
  Buta R., Corwin H.~G., Odewahn S.~C., \ 2007, The de Vaucouleurs Atlas of Galaxies,
  Cambridge Univ. Press
\bibitem[Buta \& Crocker(1991)]{buta1991}
  Buta R., Crocker D.~A., \ 1991, AJ, 102, 1715
\bibitem[Byrd et al.(1994)]{byrd1994}
  Byrd G., Rautiainen P., Salo H., Buta R., Crocker D.~A., \ 1994, AJ, 108, 476
\bibitem[Brand \& Blitz(1993)]{brand1993}
  Brand J., Blitz L., \ 1993, A\&A, 275, 67
\bibitem[Cabrera-Lavers et al.(2007)]
  {cabrera-lavers2007} Cabrera-Lavers A., Hammersley  P.~L.,
  Gonzalez-Fernandez C., Lopez-Corredoira M., Garzon F.,
  Mahoney T.~J.,\ 2007,  A\&A, 465, 825
\bibitem[Cabrera-Lavers et
al.(2008)]{cabrera-lavers2008} Cabrera-Lavers, A., Gonz{\'a}lez-Fern{\'a}ndez, C., Garz{\'o}n, F., Hammersley, P.~L., \& L{\'o}pez-Corredoira, M.\ 2008, \aap, 491, 781
\bibitem[Churchwell et al.(2009)]{churchwell2009} Churchwell, E., et
al.\ 2009, PASP, 121, 213
\bibitem[Clemems(1985)]{clemens1985}
  Clemens D.~P., \ 1985, ApJ, 295, 422
\bibitem[Contopoulos \& Grosbol(1989)]{contopoulos1989} Contopoulos, G., \& Grosbol, P.\ 1989, \aapr, 1, 261 
\bibitem[Contopoulos \& Papayannopoulos(1980)]{contopoulos1980}
  Contopoulos G., Papayannopoulos Th.,\ 1980, A\&A, 92, 33
\bibitem[Dambis et al.(1995)]{dambis1995}
  Dambis A.~K., Mel'nik A.~M., Rastorguev A.~S.,\ 1995, Astron. Lett., 21, 291
\bibitem[Dame \& Thaddeus(2008)]{dame2008} Dame, T.~M., \& Thaddeus, P.\ 2008, ApJL, 683, L143
\bibitem[Debattista \& Sellwood(2000)]{debattista2000} Debattista, V.~P., \& Sellwood, J.~A.\ 2000, \apj, 543, 704
\bibitem[Dehnen(2000)]{dehnen2000} Dehnen, W.\ 2000, \aj, 119, 800 
\bibitem[Efremov \& Sitnik(1988)]{efremov1988}
  Efremov Y.~N., Sitnik T.~G., \ 1988, Sov. Astron. Lett., 14, 347
\bibitem[Englmaier \& Gerhard(2006)]{englmaier2006}
  Englmaier P., Gerhard O., \ 2006, CeMDA, 94, 369
\bibitem[Erwin(2005)]{erwin2005} Erwin, P.\ 2005, MNRAS, 364,
283
\bibitem[Fux(2001)]{fux2001} Fux, R.\ 2001, \aap, 373, 511 
\bibitem[Hipparcos(1997)]{hipparcos1997}
 ESA 1997, The Hipparcos and Tycho Catalogs, ESA SP-1200
\bibitem[Glushkova et al.(1998)]{glushkova1998}
  Glushkova E.~V., Dambis A.~K., Mel'nik A.~M., Rastorguev A.~S.,\ 1998, A\&A, 329, 514
\bibitem[Habing et al.(2006)]{habing2006}
  Habing H.~J., Sevenster M.~N., Messineo M., van de Ven G.,
  Kuijken K., \ 2006, A\&A, 458, 151
\bibitem[\protect\citeauthoryear{Kalnajs}{1991}]{kalnajs1991} Kalnajs A.~J., \ 1991, in Sundelius B., ed., Dynamics of Disc Galaxies. G\"oteborgs Univ., G\"oteborg, p. 323
\bibitem[Kuijken(1996)]{kuijken1996}
  Kuijken K.,\ 1996, in Blitz L., Teuben P., eds, ASP Conf. Ser. Vol. 91,
  Unsolved problems of the Milky Way. Kluwer, Dordrecht, p. 504
\bibitem[van Leeuwen(2007)]{vanleeuwen2007}
 van Leeuwen, F. 2007, \aap, 474, 653
\bibitem[L{\'e}pine et al.(2001)]{lepine2001} L{\'e}pine, 
J.~R.~D., Mishurov, Y.~N., \& Dedikov, S.~Y.\ 2001, \apj, 546, 234
\bibitem[Masset \& Tagger(1997)]{masset1997} Masset, F., \& Tagger,
  M.\ 1997, A\&A, 322, 442
\bibitem[Mel'nik et al.(2001)]{melnik2001}
  Mel'nik A.~M., Dambis, A.~K., Rastorguev, A.~S.,\ 2001, Astron. Lett., 27, 521
\bibitem[Mel'nik \& Dambis(2009)]{melnikdambis2009}
 Mel'nik A.~M.,  Dambis A.~K.,\  2009, \mnras, 400, 518
\bibitem[Mel'nik \& Rautiainen(2009)]{melnikrautiainen2009}
 Mel'nik A.~M., Rautiainen  P.,\ 2009, Astron. Lett., 35, 609
 (Paper I)
\bibitem[Minchev et al.(2009)]{minchev2009} Minchev, I., Boily, C., 
Siebert, A., \& Bienayme, O.\ 2009, arXiv:0909.3516
\bibitem[Pont(1994)]{pont1994}
  Pont F., Mayor M., Burki G., \ 1994, A\&A, 285, 415
\bibitem[Rastorguev et al.(1994)]{rastorguev1994}
  Rastorguev A.~S., Pavlovskaya E.~D., Durlevich O.~V.,
  Filippova A.~A.,  \ 1994, Astron. Lett., 20, 591
\bibitem[Rautiainen \& Salo(1999)]{rautiainen1999}
  Rautiainen P., Salo H., \ 1999, A\&A, 348, 737
\bibitem[Rautiainen \& Salo(2000)]{rautiainen2000}
  Rautiainen P., Salo H., \ 2000, A\&A, 362, 465
\bibitem[Rautiainen et al.(2008)]{rautiainen2008} Rautiainen, P.,
Salo, H., \& Laurikainen, E.\ 2008, MNRAS, 388, 1803
\bibitem[Rodriguez-Fernandez
\& Combes(2008)]{rodriguez-fernandez2008} Rodriguez-Fernandez, N.~J., \& Combes, F.\ 2008, A\&A, 489, 115
\bibitem[Romero-G{\'o}mez et al.(2007)]
  {romerogomez2007} Romero-G{\'o}mez M., Athanassoula E., Masdemont J.~J.,
   Garc{\'{\i}}a-G{\'o}mez C., \ 2007, A\&A, 472, 63
\bibitem[Russeil(2003)]{russeil2003}
  Russeil D.,\ 2003, A\&A, 397, 133
\bibitem[\protect\citeauthoryear{Salo}{1991}]{salo1991}
  Salo H.\ 1991, A\&A, 243, 118
\bibitem[\protect\citeauthoryear {Salo et al.}{1999}]{salo1999}
  Salo H., Rautiainen P., Buta R., Purcell G. B, Lewis M.,
  Crocker D. A., Laurikainen E. 1999, AJ, 117, 792
\bibitem[Salo \& Laurikainen(2000)]{salo2000} Salo, H., \& Laurikainen, E.\ 2000, \mnras, 319, 377 
\bibitem[Schwarz(1981)]{schwarz1981}
  Schwarz M.~P.,\ 1981, ApJ, 247, 77
\bibitem[Sellwood \& Sparke(1988)]{sellwood1988} Sellwood, J.~A., \&
  Sparke, L.~S.\ 1988, MNRAS, 231, 25P
\bibitem[Treuthardt et al.(2008)]{treuthardt2008} Treuthardt, P.,
Salo, H., Rautiainen, P., \& Buta, R.\ 2008, AJ, 136, 300
\bibitem[\protect\citeauthoryear{Vall\'ee}{2005}]{vallee2005}
  Vall\'ee  J.~P., \ 2005, AJ, 130, 569
\bibitem[Vall{\'e}e(2008)]{vallee2008} Vall{\'e}e, J.~P.\ 2008,
AJ, 135, 1301
\bibitem[Weiner \& Sellwood(1999)]
{weiner1999}  Weiner B.~J.,  Sellwood J.~A., \ 1999, ApJ.  524,
112

\end{thebibliography}
\end{document}